\documentclass[aip,reprint, apl,twocolumn, amsmath, amssymb,floatfix]{revtex4-1}
\usepackage{graphicx}% Include figure files 
\usepackage{dcolumn}% Align table columns on decimal point
\usepackage{bm}% bold math

\begin{document}

\title{Stabilizing shallow color centers in diamond created by nitrogen delta-doping using SF$_6$ plasma treatment}

\author{Christian Osterkamp$^1$, Johannes Lang$^1$, Jochen Scharpf}
\affiliation{Institut f{\"u}r Quantenoptik, Universit{\"a}t Ulm, Albert Einstein Allee 11, Ulm 89081, Germany.}
\author{Christoph M{\"u}ller$^1$, Liam Paul McGuinness}
\affiliation{Institut f{\"u}r Quantenoptik, Universit{\"a}t Ulm, Albert Einstein Allee 11, Ulm 89081, Germany.}

\author{Thomas Diemant$^2$, R. J{\"u}rgen Behm}
\affiliation{Institut f{\"u}r Oberfl{\"a}chenchemie und Katalyse, Universit{\"a}t Ulm, Albert-Einstein-Allee 47, Ulm 89081, Germany. }
\author{Boris Naydenov$^1$}
\email{boris.naydenov@uni-ulm.de.}
\author{Fedor Jelezko$^1$}

\begin{abstract}
Here we report the fabrication of stable, shallow ($< 5$\,nm) nitrogen-vacancy (NV) centers in diamond by nitrogen delta doping at the last stage of the chemical vapor deposition (CVD) growth process. The NVs are stabilized after treating the diamond in SF$_6$ plasma, otherwise the color centers are not observed, suggesting a strong influence from the surface. X-Ray photoelectron spectroscopy measurements show the presence of only fluorine atoms on the surface, in contrast to previous studies, and suggests very good surface coverage. We managed to detect hydrogen nuclear magnetic resonance signal from protons in the immersion oil, revealing a depth of the NVs of about 5\,nm.
\end{abstract}
\maketitle

\begin{figure}[htp]
\centering
\includegraphics[scale=0.7]{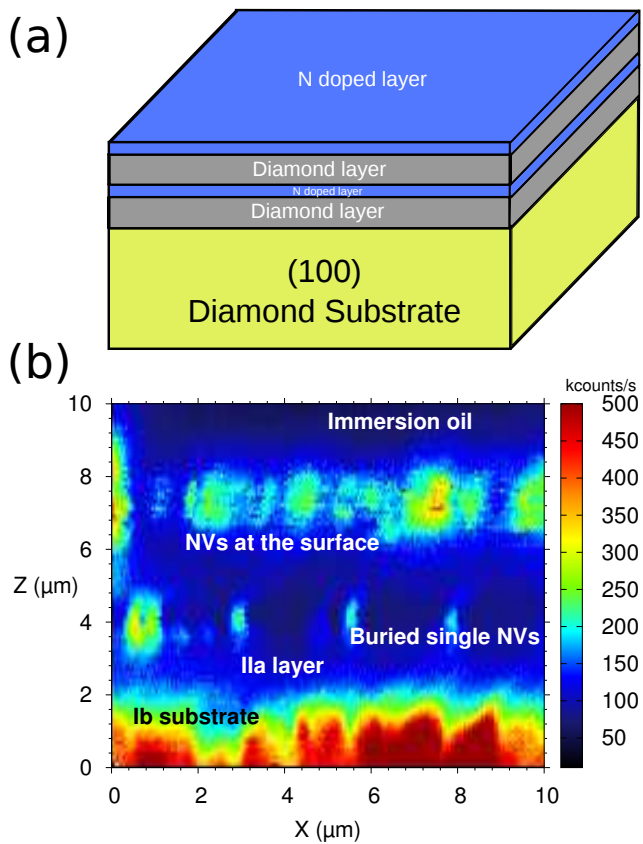}
\caption{(a) Schematic pictures of the CVD grown diamond layers. (b) Confocal image showing a cross section of the type IIa diamond layers grown on a Ib substrate. The fluorescence of single NVs in the delta-doped layers at the surface and few micron deep is shown.}
\label{Confocal}
\end{figure}

\begin{figure}[htp]
\centering
\includegraphics[scale=0.68]{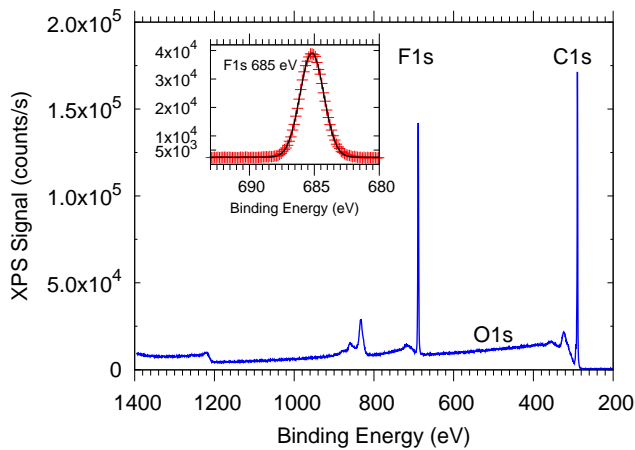}
\caption{XPS of a diamond surface treated with SF$_6$ plasma. No oxygen peak is observed, in contrary to a CF$_4$ plasma treatment \cite{Wind09,Hu13,Osterkamp13,Denisenko10}. (inset) A zoom-in of the peak at 685 eV, corresponding to F1s. The black line is a Gaussian fit to the data.}
\label{XPS}
\end{figure}

\begin{figure}[htp]
\centering
\includegraphics[scale=0.68]{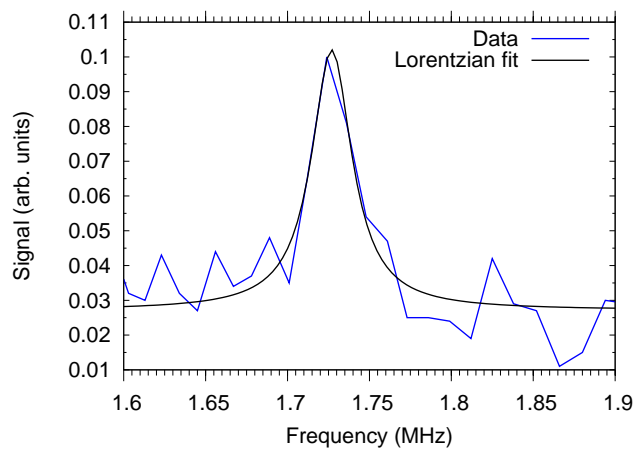}
\caption{Proton NMR spectrum of immersion oil on the diamond surface showing the proton resonance line at 1.727\,MHz, as expected. The signal strength corresponds to a depth of the NV center of about 5\,nm.}
\label{Proton_Spectrum}
\end{figure}

\begin{figure}[htp]
\centering
\includegraphics[scale=0.68]{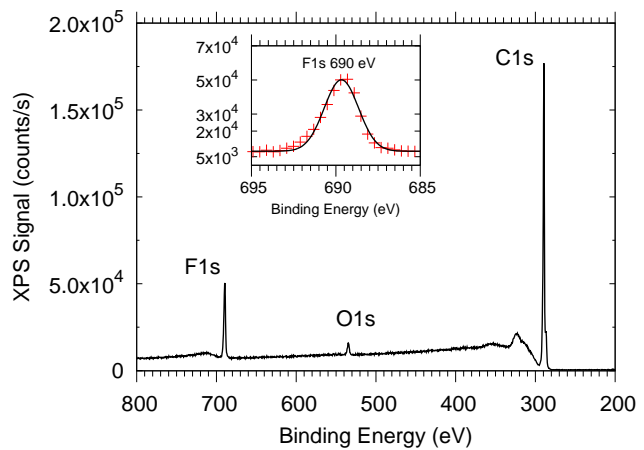}
\caption{X-Ray photoelectron spectroscopy of a diamond sample treated with SF$_6$ plasma and then boiled in the three acid mixture. A peak related to oxygen appears at 530 eV, as expected. The fluorine peak is still observed, but it is shifted by 5\,eV.}
\label{XPS_Acid}
\end{figure}

Single color defects in diamond and especially the negatively charged nitrogen-vacancy (NV) centers have unique properties making them very good candidates for single photon sources \citep{Aharonovich11}, nano-scale magnetic \cite{Gopi08, Maze08} and electric \cite{Dolde11} field sensors and quantum bits (qubits) \cite{Hanson14}. The NV can be observed at a single site level even at ambient conditions due to its strong optical transition. The fluorescence intensity depends on the electron spin state of the triplet ground state, allowing single electron spin measurements \cite{Gruber97} using optically detected magnetic resonance (ODMR). Recently, NVs situated close to the diamond surface became very popular as they can be used to detect single external electron \cite{Grotz11} and nuclear \cite{Mueller14} spins with wide applications in molecular biology and even as a quantum simulator \cite{Cai13}. Usually the shallow NVs are fabricated by low energy ($<$~5\,keV) nitrogen ion implantation, which allows to position these centers within the crystal with nanometer precision. Unfortunately this production process is limited by the reduced creation yield \cite{Pezzagna10} as well as the decrease of charge-state stability \cite{Gaebel06} and coherence time T$_2$ of the implanted NVs \cite{Naydenov10a}. Color centers produced during the chemical vapor deposition (CVD) growth process are stable and have very long T$_2$ times, which is only limited by the presence of the remaining $^{13}$C nuclear spins (nuclear spin $I=1/2$, natural abundance 1.1\,\%) \cite{Nori09} and can be increased to few milliseconds in $^{12}$C enriched diamonds \cite{Gopi09}. However, these centers are buried deep in the crystal (few micrometers) so they cannot be used for sensing of other spins on the surface as the magnetic dipole-dipole interaction decays as the third power of the distance to the surface and is therefore negligible after few tens of nanometers. To overcome this problem, the method of delta-doping can be implemented where a flow of nitrogen gas is let into the reactor chamber during the CVD growth. Depending on which stage of the growth process this is done, NVs at different positions into the crystal can be created. With this technique NVs as shallow as few nanometers can be fabricated \cite{Awschalom12,Bleszynski14,Degen13}, which show better properties (stability and long T$_2$, limited by the surface spin bath \cite{Bleszynski14}) compared to implanted ones.\\
In this Letter we report on the production of shallow NV centers using nitrogen delta doping. The main difference in the current work, compared to previous reports \cite{Awschalom12,Bleszynski14,Degen13} is that the NVs are not stable if the surface is not terminated properly. These centers could only be detected, if the surface was terminated with fluorine atoms.\\
We used commercial diamonds as substrates (E6 and Sumitomo). They were prepared for the growth process by exposing them to several acid treatments to ensure not to contaminate the reactor chamber with impurities as silicon, copper or gold. The samples are treated subsequently with: piranha acid, chromium sulfuric acid, aqua regia, ammonia-hydrogen peroxide, potassium hydroxide and finally piranha acid. They are then mounted in the reactor and exposed to a hydrogen plasma for 5~minutes to allow steady-state plasma conditions as the forward and reverse power matching, the temperature and the pressure in the CVD chamber. Furthermore the diamond surface is then hydrogen terminated and the growth process starts when methane is added. Details on the CVD growth mechanism are described in \cite{Skolov94}. First we grow a type IIa diamond layer of about 5\,$\mu$m thickness using Plasma Enhanced CVD (PECVD) on a type Ib substrate provided by Sumitomo Electric Industries, Ltd. The layers have been produced in a home built PECVD reactor at temperature $t=750$\,$^{\circ}$C using a mixture of methane (purity 99.9995\,\%, flow of 1\,sccm) and hydrogen (flow of 200\,sccm) at pressure of $p=20$\,mbar. The microwave frequency and power were set to 2.45\,GHz and 1.2\,kW respectively. The growth rate was about 200\,nm/h, calibrated by the weight difference of the samples before and after the growth process. Confocal microscopy measurements show very few NVs, suggesting very low nitrogen concentration, probably in the ppb range (see figure~\ref{Confocal}).\\
%It is interesting to note, that the methane (99.9995 \%) was not specifically purified from nitrogen.
After being able to grow reproducibly clean diamond layers, we performed the delta doping procedure. During this procedure nitrogen gas (flow of 5\,sccm) was added into the reactor chamber for 5~minutes and afterward the doped layer was overgrown with a further diamond layer of several micron thickness. In this buried doped layer, single NVs with a concentration of about 1 per 5 $\mu$m$^2$ were found. The electron spin coherence time $T_2$ of these centers was measured to be few hundreds of $\mu$s (data not shown), limited only by the $^{13}$C nuclear spins and excluding the presence of other paramagnetic defects as reported earlier \cite{Nori09}. Additional growth on IIa substrates (electronic grade from Element Six, Ltd), did not show any observable difference in the properties of the grown layers.\\
Nitrogen delta-doping is again repeated at the same conditions as for the deep NVs and immediately after this the CVD process is stopped. Since the carrier gas in the plasma chamber is hydrogen, the surface after the growth process is hydrogen terminated. The latter is known to alter the charge state of shallow ($<$~10\,nm)  NV centers and leads to disappearance of the fluorescence \cite{Hauf11}. Therefore the diamond is boiled in an mixture of three acids (nitric, sulfluric and perchloric acid, volume ratio 1:1:1) at $t=180$\,$^{\circ}$C to change the surface termination to oxygen which shifts the Fermi level at the surface and makes the NV$^-$ charge state the dominant one. But even after this procedure we could not detect NVs which were close enough to the surface to detect a Nuclear Magnetic Resonance (NMR) signal from the hydrogen nuclei contained in the immersion oil (see below).\\
In our previous work \cite{Osterkamp13} and in a related work \cite{Hu13} it has been shown, that very shallow ($<$~5\,nm) implanted NVs are stabilized if the diamond surface is treated with oxygen or CF$_4$ plasma. Here we report a similar effect, namely NVs produced during the CVD process and close to the surface are not stable neither if the samples are measured directly after the growth nor after boiling them in the three acid mixture. Only few centers can be observed and they show blinking - the fluorescence disappears and appears on a time scale of few seconds. Additionally the fluorescence spectrum of the very few stable centers is similar to the spectrum of the neutrally charged NV$^0$ and ODMR signal was not detectable.\\
For the samples in this study we treated the surface with plasma, but we decided against the known CF$_4$ procedure as according to some reports \cite{Denisenko10,Wind09} the latter builds a thin layer (2-5\,nanometers) of a polymer similar to Polytetrafluoroethylene (PTFE) on the top of the diamond surface. Furthermore, the surface fluorine coverage is not complete after this process. Instead we chose the SF$_6$ plasma, commonly used for diamond etching \cite{Goto01}, where we expected to obtain a smooth surface \cite{Nebel14} with better fluorine coverage, without any polymer layer or unwanted oxygen groups. We applied the procedure reported in \cite{Osterkamp13}. First we hydrogenate the diamond surface by exposing the sample in hydrogen plasma for several minutes at temperature $t=750$\,$^{\circ}$C and pressure $p=20\,$mbar in the  CVD reactor used for the growth. After that the diamond is treated with SF$_6$ (gas flow 100\,sccm) plasma (Oxford Plasma Lab 100) for about 4\,minutes with power of the Inductively Coupled Plasma (ICP) of 500\,W. This exposure time does not lead to observable etching of the diamond, but nevertheless the surface is covered with fluorine. X-Ray photo-electron spectroscopy (XPS) measurements shown in figure~\ref{XPS} confirm the presence of fluorine on the surface.
The fluorine F1s peak appears at 685\,eV and it is shifted by 3\,eV compared to the position of the CF$_4$ treated samples \cite{Osterkamp13}. We do not observe any oxygen peak, expected at around 530\,eV, suggesting a good fluorine surface coverage. Nevertheless, from our data we cannot conclude that there is a fluorine monolayer on the surface. In these samples we observed an increased number (comparable with the deep nitrogen doped layer) of stable NVs, no blinking and no photochromism. Additionally the NVs showed good ($>$~\,20\,\%) ODMR contrast. Our results contradict a recent study, where a significantly reduced number of NVs close to a fluorine terminated surface (by using CF$_4$ plasma) is reported \cite{Degen13}. In our experiments we do not observe any change of the stability after applying the SF$_6$ plasma process. On the contrary, we observe the opposite effect, an increased NV$^-$ concentration (no NV$^0$), in agreement with previous reports on bulk diamonds \cite{Osterkamp13, Hu13} and in nanodiamonds \cite{Aharonovich14}. The effect reported in \cite{Degen13} could be related to etching of the  diamond surface. Kaviani et al. have performed Density Functional Theory (DFT) calculations for NVs situated one nm below a diamond surface with various chemical groups \cite{Gali14}. In the case of fluorine the authors predict the creation of localized surface states with energies close to the conduction band minimum. They can trap the electron from a NV$^{-}$ and change it to NV$^{0}$. However, this process happens on a time scale (few nanoseconds), much shorter compared to our measurements (milliseconds), and is probably not relevant. Additionally our observations are supported by a band banding model \cite{Hauf11,Pakes13}.\\
T$_2$ measurements revealed a rather short average coherence time of about T$_2 = 4\,\pm1\,\mu$s, typical for implanted NV. This  value could be explained by a strong influence from paramagnetic defects (e.g. dangling bonds) on the diamond surface.\\
In order to determine the depth of the NVs, we measured the NMR signal originating from hydrogen atoms contained in the immersion oil. With this method the distance of NVs to the diamond surface can be determined \cite{Staudacher13,Degen13}. In our samples about 90\,\% of the measured centers showed an NMR signal, a typical spectrum is shown in figure~\ref{Proton_Spectrum}.
The area of the peak is proportional to the magnitude of the magnetic field $B_{\mathrm{H}}$ created by statistical polarization of the external proton spins. The latter can be directly related to the depth of the NV as described in \cite{Staudacher13}. Our measurements revealed an average depth of the centers of about 5\,nm.\\
In order to confirm that the stabilization of the shallow NVs indeed originates from the fluorine on the surface, we tried to remove the fluorine by treating the samples in the three acid mixture (see above). To our surprise, the XPS measurements still showed the presence of fluorine even after boiling the diamonds for many hours, see figure~\ref{XPS_Acid}.
Nevertheless, an oxygen O1s peak appears at the expected position (530\,eV) and the intensity of the F1s peak is significantly reduced. The latter is shifted by about 5\,eV (690\,eV), compared to the SF$_6$ (685\,eV) treated surface, while the carbon C1s peak remained at the same position. Similar position has been previously reported for surfaces treated with CF$_4$/O$_2$ plasma. This result suggests that the fluorine has a different chemical environment, changed by the acid treatment, e.g it might be that a PTFE type of polymer is build on the surface. After this procedure, we were not able to observe stable NVs in the delta-doped layers. This is a strong indication that the SF$_6$ treatment indeed stabilizes the NVs.\\
In conclusion we have demonstrated the creation of NV centers in ultra pure diamond layers by nitrogen delta doping. These centers show a coherence time limited only by the dynamics of the $^{13}$C nuclear spins bath. By controlling the doping process we can produce NVs as close as 5\,nm to the diamond surface. The distance is determined by measuring NMR of protons from the immersion oil. Additionally we show that the NVs need to be stabilized by terminating the surface with fluorine atoms, otherwise we could not find centers in these layers.We believe that the results reported here will be useful for designing sensors based on shallow defect centers in diamond.

\begin{acknowledgments}
This work has been supported by DFG (SPP1601/1, SFB TR21, FOR1493), Volkswagenstiftung, EU (STREP Project DIADEMS, ERC Synergy Grant BioQ and EQuaM). BN is grateful to the Bundesministerium f\"{u}r Bildung und Forschung (BMBF) for the  ARCHES award.
\end{acknowledgments}

%\bibliography{Delta_Doping}
%
\end{document}